\documentstyle[twocolumn,prc,aps,psfig]{revtex}
\begin{document}\hbadness=10000
\twocolumn[\hsize\textwidth\columnwidth\hsize\csname %
@twocolumnfalse\endcsname
\title{$B_c$ Meson Production in Nuclear Collisions at RHIC
}
\author{Martin Schroedter, Robert L. Thews, and Johann Rafelski}
\address{
Department of Physics, University of Arizona, Tucson, AZ 85721.
}
\date{April 4, 2000}
\maketitle
\begin{abstract}
We study quantitatively  the formation and 
 evolution of $\bar b c,\,b\bar c$ bound states
in a  
space-time domain of deconfined quarks and gluons 
(quark-gluon plasma, QGP).  
At the Relativistic 
Heavy Ion Collider (RHIC), one expects for the first time 
that typical central collisions will result in multiple pairs
of heavy (in this case charmed) quarks.  This provides a new
mechanism for the formation of heavy quarkonia which depends on
the properties of the deconfined region.
We find typical enhancements of about 500-fold 
for the  $\bar b c,\,b\bar c$
production yields over expectations from the 
elementary coherent hadronic 
$B_c$-meson production scenario.  The final population of bound 
states may serve as a probe of the
plasma phase parameters.\\

\end{abstract}
]
\begin{narrowtext}
\section{Introduction}\label{intro}
The recent observation of candidate $B_c$-meson events by the
CDF collaboration \cite{CDF98} yields measurements for the ground
state mass and lifetime which are consistent with expectations
from nonrelativistic potential models \cite{Ger95}. 
This system of states is expected to have properties intermediate
between the $J/\psi$ and $\Upsilon$ systems (except for its longer
lifetime due to the absence of quark annihilation processes).  Thus
it is natural to investigate the fate of such states when produced
in a deconfined environment, where suppression relative to
``expected" yields may serve as a signature for the existence
of deconfinement.  

We have previously reported some preliminary work toward this end
\cite{FTR99,TSR99}.  Our calculations of the expected yield of
$B_c$ at RHIC utilize estimates of the coherent production of
both a $b\bar b$ and $c\bar c$ pair in the same initial hard
perturbative QCD process, followed by hadronization into the
$B_c$ states.  This process is of order $\alpha_s^4$, and falls quite
rapidly with decreasing energy.  The resulting prediction at
RHIC energies for the bound-state fraction relative to 
initial b-quark production is in the range $10^{-4}\,-\,10^{-5}$  
\cite{KR98}.  Combining this result with the expected
yield of b-quarks, we find that even at design luminosity there
will be at most a handful of $B_c$-mesons produced at RHIC
which decay into observable final states  
(all  scalar, vector, excited, particle/antiparticle
$\bar b c$ bound states will
be called $B_c$-mesons or simply $B_c$, except when otherwise noted).
Thus a
scenario of production via calculable hard QCD interactions
followed by suppression in a deconfined medium will be irrelevant
for RHIC parameters.

What we now present is a new production mechanism 
which itself depends on the existence of a deconfined state.
The new mechanism becomes relevant for the first time at RHIC
energies, since typical central collision events will have
multiple pairs of initially-produced charm quarks.
Estimates using perturbative QCD to calculate the initial
production of heavy quarks predict approximately
10 $c\bar c$ pairs in each central event in Au-Au collisions
at $\sqrt{s}$ = 200 GeV \cite{HPC95}.  Due to the larger mass, 
only about
1 in 20 central events will produce a $b\bar b$ pair.

Consider the events in which there is a
single $b\bar b$ pair, along with the expected 10 $c\bar c$ pairs.
If a region of deconfined matter is subsequently
produced in the space-time region encompassing the heavy quarks, 
a b-quark will be able to ``find" any of the initially-produced
charm quarks with which to produce the final $B_c$ bound state.  
More generally, 
our study shows that the formation of quarkonium states
within a deconfined space-time domain
resulting from the interaction of mobile heavy quarks 
offers a very  interesting signature
for deconfinement.  

In Sec.\,\ref{chem} we study quantitatively the probability that 
in the deconfined quark-gluon plasma (QGP) phase the mobile
$c,\,\bar c$-quarks can seek, find and bind to a $\bar b,\, b$ 
quark produced in the same event.  
We also consider 
the dynamical evolution of these
heavy quark bound states in the deconfined phase. 
In Sec.\,\ref{ratechem} we establish the kinetic model for
the evolution in time of the  $\bar b c,\,b \bar c$ bound state
population. The required cross sections are presented in Sec.\,\ref{sigma},
and the formation and dissociation reaction rates obtained.

To calculate the final $B_c$ yield, one needs an estimate of
the charm quark {\it density} during the deconfined phase.
We therefore consider in Sec.\,\ref{Tevol} a generic model 
for the expansion and cooling of the QGP. 
This also enables the comparison of direct (initial)
and microscopic thermal charm production in Sec.\,\ref{charm}.
We confirm that at RHIC energy charm is produced
primarily in the initial parton interactions, with 
only a minor contribution arising from the thermal 
plasma processes. We also show that the charm density at 
hadronization remains significantly in excess of that which would
be present if full chemical equilibrium were reached.

We  then present  in Sec.\,\ref{results} a study  of the 
influence of the model parameters on
the pattern of $B_c$ production at RHIC. We present in 
Sec.\,\ref{Bcyield} the fractional yield per initial bottom quark pair
in QGP. 
We find that the final yield is most sensitive  to the initial charm density,
and there is very little dependence 
on the initial temperature  
of the dense deconfined state.  
In contrast, we show
in Sec.\,\ref{2sBcyield} that the relative yield 
of the first radially excited state to the ground state, 
$B_c(2S)/B_c(1S)$ is significantly more 
dependent on the initial temperature in the plasma phase.

\section{Chemical kinetics of quarkonium abundance}\label{chem}
In our scenario, the $b\bar c$ and $\bar b c$ states formed in the QGP will 
be subject to collisions with gluons, which
will dissociate them into their constituent quarks.  This mechanism
can be thought of as 
the dynamic counterpart of the plasma screening scenario, in which
the color-confinement force is screened away in the hot dense plasma 
\cite{Kha94,Kha96}.  We do not distinguish between the
vacuum and plasma values for the mass and binding energy of the 
1S ground states.
Both are significantly larger than typical temperatures expected 
at RHIC.  Estimates for their behavior as a function of
screening mass have been made \cite{FULCHER}, and indicate that
such an approximation is reasonable.  Thus we no longer distinguish
between the plasma bound states and the physical $B_c$ mesons which can be
observed after hadronization.

The primary formation mechanism is just the inverse of
the breakup reaction, in which unbound heavy quarks are captured in the
bound states, emitting a color octet gluon.  The competition
between the rates of these reactions integrated over the lifetime of
the QGP then determines the final $B_c$ populations.  Note that in
this scenario it is impossible to separate the formation process
from the breakup (suppression) process.  Both processes occur
simultaneously, in contrast to the situation in which
the formation only occurs at the initial times before the QGP is 
present.  (Of course, one can include this case also in our scenario
by adjusting the initial conditions.  However, for the case of
$B_c$ initial production at RHIC, we have already shown that this
possibility is negligible.) 
A number of other reactions involving b-quarks are possible in the QGP, but
the rates are much smaller than those above.  For example, formation of
bound states with light quarks, $B_s, B_u, B_d$, are not possible
at the high initial temperatures expected at RHIC since their lower
binding energies prevents them from existing in a hot QGP, or 
equivalently they are ionized on very short time scales.
They will be formed predominantly at hadronization, but
this process is too late to affect the final $B_c$ population.
We also neglect the decrease in b-quark population due to $b\bar b$ 
annihilation into light quarks or formation of $\Upsilon$ states. 
Both of these processes depend quadratically on the b-quark
densities, and proceed too slowly or too rarely, respectively, to
be significant.  We also neglect the breakup cross section 
due to collisions with light quarks, since their population in the
QGP is expected to be very much suppressed relative to gluons
as compared to chemical equilibrium values during the early
times when the breakup reaction rate has its most significant
effect \cite{NONEQ}.\\ 

\subsection{Chemical rate equations}\label{ratechem}
The abundance
of bottom and charm quarks and their bound states is thus governed
by very simple master (population) equations involving only
two reactions. Specifically, the 
formation reaction F 
$$b+\bar c \rightarrow B_c+g\,,$$
and the dissociation reaction D
$$ B_c+g \rightarrow b+\bar c\,,$$
and similar equations for the conjugate states 
determine the time evolution of the number 
of  bound states in the deconfined
region, $N_{B_c}$.
\begin{equation}\label{eqNBc}
\frac{dN_{B_c}}{d\tau}=
  \lambda_{\mathrm{F}} N_b\, \rho_{\bar c } - 
    \lambda_{\mathrm{D}} N_{B_c}\, \rho_g\,.
\end{equation}
Since the total number of b-quarks does not change
under the assumptions above,
the rate of change in the number of unbound $b$ quarks $N_b$ is just
the negative of that for the number of bound state $B_c$ mesons :
\begin{equation}\label{eqNb}
\frac{dN_b}{d\tau}=
     \lambda_{\mathrm{D}} N_{B_c} \,\rho_g - 
            \lambda_{\mathrm{F}} N_b\, \rho_{\bar{c}}\,.
\end{equation}
In Eqs.\,(\ref{eqNBc},\ref{eqNb}),  $\tau$ is the proper time
in a small volume cell, $\rho_i$ denotes the 
number density $[L^{-3}]$ of species {\it i} and 
the reactivity $\lambda$ $\left[L^3/\mathrm{time}\right]$ is 
the momentum  distribution averaged reaction rate:
\begin{equation}\label{eqReact}
\lambda\equiv \langle \sigma v_{\mathrm{rel}} \rangle 
  =\frac{
   \int \int d^3p_1 d^3p_2  f_1(p_1)f_2(p_2)\,
  \sigma({\sqrt{s}})v_{\mathrm{rel}}}
  {
   \int \int d^3p_1 d^3p_2  f_1(p_1)f_2(p_2)
  }\,,
\end{equation}
Note that even though $\lambda$ and $\rho$ are not Lorentz invariant,
their product, the rate of particle production, is invariant. Thus, we evaluate
the rates $\rho_i\, \lambda_{\mathrm{D/F}}$
in the volume cell frame, 
where the (local) densities are given by external inputs.

We study here the deconfined period during which
parton distributions have kinetically 
(but not necessarily chemically) equilibrated.  We thus 
consider an  isotropic medium of mobile quarks and gluons 
with the momentum distribution function $f_i$
for particle {\it i} given by the thermal
equilibrium distribution $f=(e^{E/T}\pm1)^{-1}$.
$\sigma$ is the spin- and color-averaged
total cross section for the relevant reaction
which depends only on $s = (p_1+p_2)^2$, and 
$v_{\mathrm{rel}}$ is the relative
speed between the two reacting particles. 
Except where
 otherwise indicated $\rho_g$ is assumed to be in chemical equilibrium. 
 $\rho_{\bar{c}}$ will be
determined by its own kinetic equation described in
Sec.\,\ref{charm}.                                    
The spatial distribution of charm and bottom
quarks will be taken to be uniform, an approximation which indicates
the accuracy to which we will pursue our following calculations.

\subsection{In-plasma cross section and reaction rates}\label{sigma}
We now need to estimate the cross sections involved in the formation
and breakup reactions.  For these purposes, we first utilize a
derivation based on the interaction of a gluon field with the
color dipole moment of a nonrelativistic heavy quarkonium state.
It is implemented via an operator product expansion technique
\cite{OPE}, 
and has been applied to the $J/\psi$ breakup rates in a QGP \cite{Kha96}.
We generalize this result to any heavy quarkonium 1S state with arbitrary
flavor content, so that the spin and color averaged dissociation  
cross section is written
\begin{equation}\label{eqSigmaB}
\sigma_{\mathrm{D}} =
 \frac{2\pi }{3} \Big(\frac{32}{3}\Big)^2 
     \Big(\frac{2\mu }{\epsilon }\Big)^{1/2}
        \frac{ k_0^{7/2}}{4 \mu^2}
             \frac{(k-k_0)^{3/2}}{k^5 }\,.
\end{equation}
Here $k$ is the gluon momentum (in the quarkonium-rest frame), 
$k_0$ is the minimum value required to 
impart the binding energy $\epsilon$ to the bound state, and $\mu$ is
the reduced mass. This form is valid if the quarkonium system has
a spatial size small compared with the inverse of $\Lambda_{QCD}$, and
its bound state spectrum is close to that in a nonrelativistic
Coulomb potential with $\epsilon$ large compared with $\Lambda_{QCD}$. 
These conditions are marginally satisfied for the $J/\psi$, and
should be somewhat better for the $B_c$ kinematics.  The form above
has also been altered to account for recoil of the finite mass system,
since in the original form the values of $\epsilon$ and $k_0$ were
identical.

We use values  $m_b$ = 5.8 GeV and
$m_c$ = 1.3 GeV which are consistent with typical potential
model fits to the spectra.  The magnitude of the 
cross section is controlled
by the geometrical factor $(4\mu^2)^{-1}$, which for these
quark masses is consistent with the size of the bound state
wave function in the same potential models.  The rate of increase
just above threshold is due to phase space and the p-wave color
dipole interaction, and it reaches a maximum value
for $B_c$ dissociation of about $1.5$ mb when $k = \frac{10}{7}\, k_0$.
For our model
calculations we use a central value of 6.3 GeV for the $B_c$ mass, 
and a binding energy of 0.84 GeV which follows 
from the hadronic open flavor B and D
meson masses.  We retain these values for our kinetic calculations,
as an approximation for physical values in the QGP.  One might
expect that both would decrease somewhat in a QGP.  Some potential
model calculations utilizing screening indicate the magnitude
of this effect is expected to be small for the $B_c$ at RHIC
conditions \cite{FULCHER}.   

Our dominant  formation cross section $\sigma_{\mathrm{F}}$ can then
be directly obtained from  
 $\sigma_{\mathrm{D}}$  
by utilizing the detailed balance relation. This is written in 
the zero-momentum (ZM) frame of two-body interactions as
\begin{equation}\label{eqDetailedBalance}
 (\sigma\, {\mathbf{p}}^{2}\, g)_{\mathrm{D}} =
 (\sigma\, {\mathbf{p}}^{2}\, g)_{\mathrm{F}}\,,
\end{equation}
where ${\mathbf{p}}_{\mathrm{D/F}}$ is the 3-momentum of the 
initial state particles for these reactions in their respective
ZM frames, and
$g_{\mathrm{D/F}}$ 
is the statistical degeneracy in the two channels.
In the present case of an unpolarized QGP, 
$g_{\mathrm{D}} = ((3+1)\cdot 1)(2\cdot 8)$, counting 
spin and color multiplicities of $B_c$ color-singlet and gluon, 
and $g_{\mathrm{F}}= (2\cdot 3)(2\cdot3)$,
counting spin and color multiplicities of the initial state 
b- and c-quarks. 
In this formalism we have included both the pseudoscalar and vector
1S $B_c$ states with the same cross sections, since in the
nonrelativistic approximation one would expect the same spatial
wave function and no spin dependence.  These two states are also
included in the no-deconfinement scenario initial production
estimates \cite{KR98}, so that a direct comparison can be made. 
Using these formation - dissociation cross sections we calculate 
the reactivities as defined by Eq.\,(\ref{eqReact}). As shown 
in Fig.\,\ref{figLambdaBF},  formation and dissociation tend to balance 
at very high temperature, where the endotherm nature of the dissociation
is negligible. At temperatures in the range 150--300 MeV the formation
reactivity dominates by about a factor 4. 
\begin{figure}[t]
\centerline{  \hspace*{-.cm}
\psfig{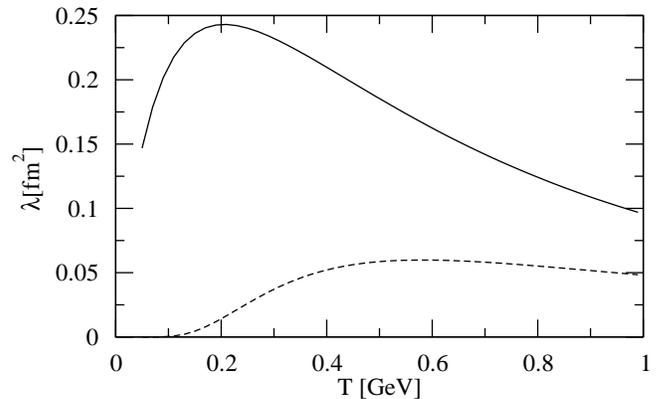}
}
\caption{ \small
 Formation  $\lambda_{\mathrm{F}}$ (solid)
 and dissociation $\lambda_{\mathrm{D}}$ (dashed)  reactivities
 as function of temperature.
\label{figLambdaBF}} 
\end{figure}

\subsection{$B_c$ dissociation time scales}\label{timescales}
From Eq.\,(\ref{eqNBc}), one sees that if the system is at a
constant temperature (or equivalently the reaction rates are sufficiently
fast), the final ratio of bound state to free quark populations is
just given by the ratio $(\lambda_{\mathrm{F}}\rho_{\bar c})/
(\lambda_{\mathrm{D}}\rho_g)$.
The relevant time scales are set by the magnitudes of either
factor in this ratio.  We have calculated the dissociation rates
of several quarkonium states under breakup by gluons with full
chemical equilibrium density, 
and the results are shown in Fig.\,\ref{figdiss}.

As expected, these rates rise quite sharply with temperature.  The
$B_c$ curve fits quite nicely between those for the $J/\psi$ and
$\Upsilon$, also as expected.  We also show the same calculation
for the $B_s$ state, although the approximations made for this cross
section have a very marginal validity in view of such a large state
with small binding energy.  For the $B_c$, one sees that in the
range of initial temperatures expected at RHIC (roughly 300 to 500
MeV), these dissociation rates imply time scales of order 1--10 fm/c.
Since the total QGP lifetimes are also in this region, it is evident
that the equilibrium solutions will not in general be reached at each
temperature, and one must solve the rate equations numerically
to obtain the final populations.

\begin{figure}[t]
\centerline{  \hspace*{-.cm}
\psfig{width=8.5cm,clip=,figure=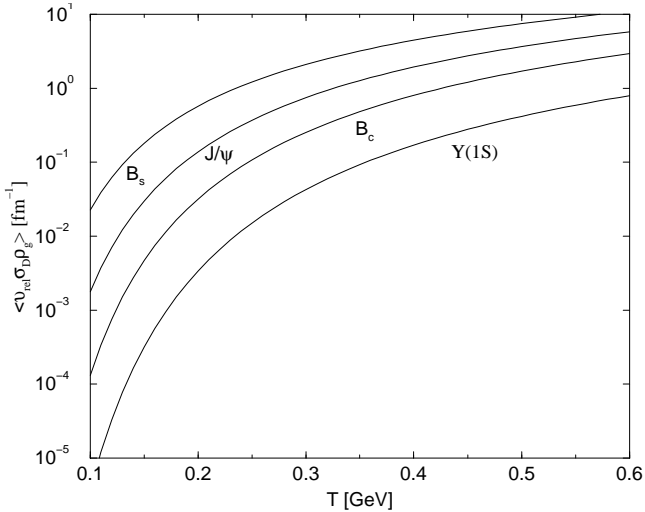}
}
\caption{ \small
Thermal QGP quarkonium dissociation rates by thermal gluons as functions
of temperature.
\label{figdiss}} 
\end{figure}

\section{Evolution of charm density in QGP}\label{evolcharm}
\subsection{Evolution of temperature}\label{Tevol}
In order to obtain the chemical evolution of quark bound state 
abundances in QGP
we need to establish a relation between plasma temperature and proper time. 
We assume that the expansion of the QGP follows an isentropic 
path \cite{Bjo83}. We utilize a generic
scenario for the proper-time dependence of the volume involving both 
longitudinal and transverse expansion, and examine the sensitivity
of our final results to variations in the parameters involved.
As a specific example we consider  
an adiabatically expanding homogeneous QGP domain 
having the shape of an ellipsoid of revolution
about the longitudinal axis with semi-major and semi-minor 
axes parameterized with an initial length $l_i$/2 = $\tau_0$
and an initial transverse
radius $r_i$. 
These are fixed at the time of QGP equilibration 
at an initial temperature $T_0$, when our formation and breakup
reactions are assumed to start. 
We will explore the range $ 0.5<l_i<2$ fm (equivalent to thermalization
times between 0.25 and 1.0 fm), and take $r_i$ = 5 fm 
for Au--Au collisions at RHIC, corresponding to 15--20\% most central
interactions. 
The longitudinal growth occupies the region between
the (almost unstopped) receding nuclei.  
For transverse growth we allow a radial expansion
at a speed $v_r$. The
speed of sound in an ideal relativistic
gas $v_r\simeq 0.58$ c is taken as a nominal value, but the effects 
of significant
variations in parameter space will be considered.
Thus  the volume evolves 
as a function of proper time $\tau$ according to:
\begin{equation}\label{eqVolume}
 V(\tau) = 
 \frac{4 \pi}{3} {(r_i+v_r\,\tau)^2\, (\frac{l_i}{2}+ \tau)}\,.
\end{equation}
Note that we have rescaled the initial proper time to zero.

This simple approach produces 
temperature vs. time profiles which appear to be very similar to
those arising in more complex studies.
These are shown in Fig.\,\ref{figTemp} for a homogeneous bulk QGP state
in a range of initial temperatures $0.3\le T_0\le 0.6$\,GeV.
Intersection with the horizontal line at freeze-out
temperature $T_f = 0.15$\,GeV indicates the QGP lifespan
between thermalization and hadronization.

\begin{figure}[t]
\centerline{  \hspace*{-.cm}
\psfig{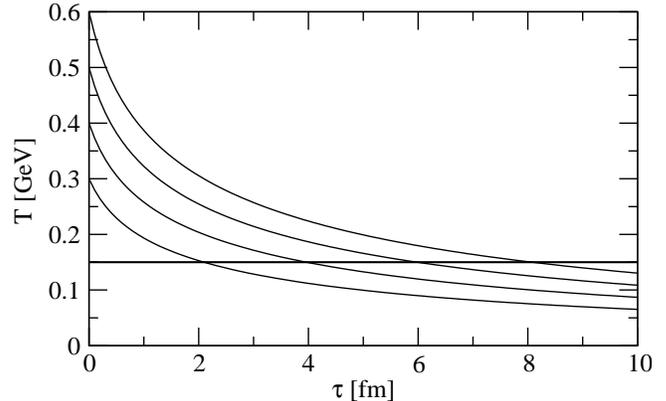}
}
\caption{ \small
 Temperature [GeV] vs. proper time [fm] for various initial temperatures
 $T_0$, with
 parameters: $l_i$ = 1 fm, $v_r$ = 0.58. The
 horizontal line marks hadronization.
\label{figTemp}}
\end{figure}

\subsection{Thermal charm production in QGP}\label{charm}
From fits to experimental nucleon-nucleon reaction  data  
it is estimated that at RHIC there
will be an average number $N_{c_0}$ = 10 of directly produced $c\,\bar c$
pairs per central collision \cite{HPC95}, and this is the standard value
for which our calculations will be carried out. However, 
additional charmed quarks could be produced 
in the QGP by collisions of gluons:
$$ g+g \to c+\bar c\,,$$
and of light quarks:
$$ q+\bar q \to c+\bar c\,.$$
The local evolution of charm density $\rho_c$ can be described in
a fashion quite similar to the model
developed for strangeness production, see e.g. \cite{Raf99}. 
Allowing for charm production and  volume dilution 
under adiabatic expansion,
the local charm density obeys: 
\begin{equation}\label{eqCharmDensity}
 \frac{d\ {{\rho }_c}}{d\ T} = 
 \frac{{A_c}[T]}{\dot{T}}\Big(1-{{\Big(\frac{{{\rho
 }_c}[T]}{\rho _{c}^{\infty }[T]}\Big)}^2}\Big)+3\ {{\rho }_c}\frac{1}{T}\,,
\end{equation}
where $\dot{T} \equiv {\mathrm{d} T}/{\mathrm{d} \tau}$\,. The
 rate of charm production is
\begin{equation}\label{eqAcT}
 A_c[T] =
 \frac{1}{2} \rho_g^{{{\infty }^2}}{\lambda^{{gg}\rightarrow c\bar{c} }}
 + \rho_q^{{{\infty }^2}}{{\lambda }^{q\bar{q} \rightarrow c\bar{c} }}\,,
\end{equation}
where $\rho^{\infty }$ denotes the density in 
thermal and chemical equilibrium.
$A_c[T]$ is shown 
in Fig.\,\ref{figAc}, obtained with running QCD 
parameters, $\alpha_s(M_Z)=0.118, m_c(1\mathrm{GeV})=1.5$\,GeV 
\cite{acta96}. 

\begin{figure}[t]
\centerline{  \hspace*{-.cm}
\psfig{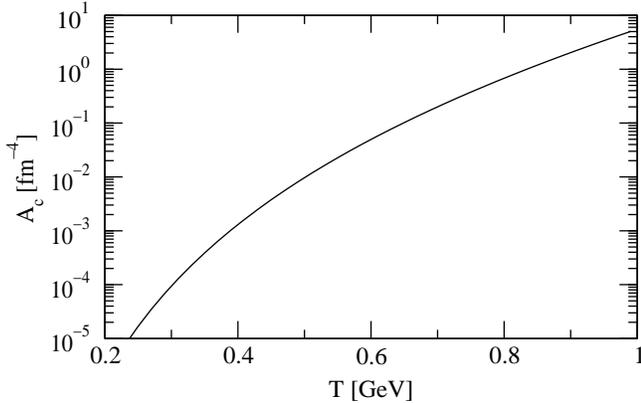}
}
\caption{ \small
(Invariant) charm production rate per unit of time and volume 
$A_c[T]$ as function of temperature $T$. 
\label{figAc}} 
\end{figure}

Integrating Eq.\,{\ref{eqCharmDensity}}, we obtain
the total number of charmed quark pairs in plasma 
at freeze-out, with initial temperature $T_0$ being a parameter, along with 
the initial longitudinal size  $l_i$\,. 
Fig.\,\ref{figCharm} shows our results for three different initial 
longitudinal sizes  $l_i = 2$\,fm (dotted),   $l_i=1$\,fm (dashed),
  and   $l_i=0.5$\,fm (solid): the  three thick up-curving lines
include both 
the 10 directly 
produced charm pairs and the QGP charm production.  
The larger the initial volume and the longer the lifespan at high 
temperature, the greater is the QGP contribution to  
charm yield. 

\begin{figure}[t]
\centerline{  \hspace*{-.cm}
\psfig{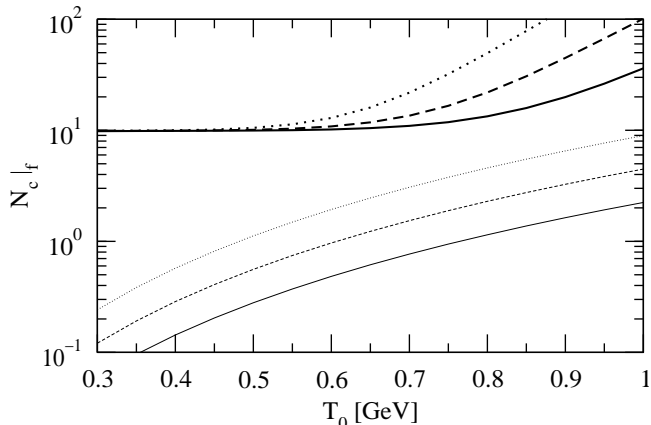}
}
\caption{ \small
 Thick lines: $N_c$ at freeze-out as function of
 initial temperature $T_0 $ [GeV]: thermally produced
 charm is combined with
 $N_{c0}$ = 10 charm pairs from direct
 initial production. Longitudinal sizes
  are $l_i$ = 2 fm (dotted), 1 fm (dashed),
  and 0.5 fm (solid); other parameters same as
  Fig.\,\protect{\ref{figTemp}}.
  Thin lines: chemical equilibrium abundance at freeze-out,
  $V\,\rho_c^{\infty }|_{T_f}$.
\label{figCharm}}
\end{figure}

For the range of initial temperatures expected at RHIC, ones sees that 
there is virtually no additional charm produced in the QGP, 
and also that
charm annihilation processes are too slow to significantly reduce the
initially-produced number of charm quark pairs.
The thin lines show for comparison the number 
of charm quarks which would
follow from a chemical equilibrium density 
at the hadron freeze-out 
temperature $T_f=0.15$ GeV for $m_c=1.5$ GeV. 
(This value depends on the initial temperature $T_0$ through its effect on
the freezeout volume.)   The chemical equilibrium
value is always significantly smaller than the corresponding 
direct + thermally produced abundance, i.e. direct initial charm
production at RHIC is predicted  
in general to significantly 
oversaturate the  statistical phase space at freeze-out. 

\section{Discussion of Results}\label{results}
\subsection{$B_c$ survival in plasma}\label{survive}
To see what temperature range contributes to breakup,
Fig.\,\ref{figBcSurvival} shows the survival
probability of one $B_c$ meson being placed in the
QGP at different initial temperatures.
At low initial temperature ($T\lesssim 350$\,MeV) the
state will most likely remain bound, while already
at $T\gtrsim 500$\,MeV the plasma will most likely
dissolve the $B_c$. One can infer from this result that
the final state population is mainly dominated by the
formation process as the plasma cools between these temperatures.  
This of course is not in exact correspondence with the actual
physical process, which involves the formation of the
bound state at any time after QGP forms at the initial temperature,
where the expansion has already produced a partial decrease
in the gluon density.

\begin{figure}[t]
\centerline{  \hspace*{-.cm}
\psfig{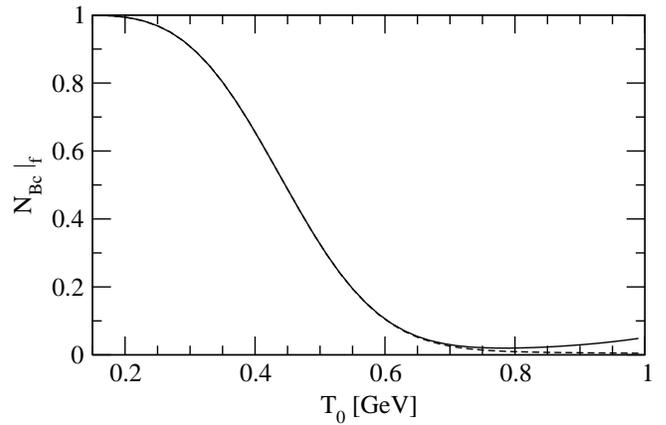}
}
\caption{ \small
 Survival probability of an initial $B_c$ as function of initial temperature $T_0
$ [GeV];
 other parameters same as Fig.\,\ref{figTemp}. Solid line: $N_c=10+$thermal
 production,
 dashed line: $N_c=10$.
\label{figBcSurvival}}
\end{figure}
Another way to interpret the results shown in Fig.\,\ref{figBcSurvival}
is that the effective collision-mediated color screening is most
effective in dissolving  
$B_c$ bound state when the average gluon energy 
$\overline E_g\simeq 3T$ exceeds the binding energy. 
Using this argument we can also easily see that
the flavor transfer reaction $\bar b s+c \rightarrow \bar bc+s$ is
unlikely, since the $\bar b s$ bound states have much smaller
binding energies and are color
screened already at all  temperatures above the hadronization
temperature.


\subsection{Fractional $B_c$ yields}\label{Bcyield}
We numerically integrate the evolution of the $B_c$ abundance 
as function of the 
temperature $T$, using our expansion model to obtain the temperature as
a function of time.  The initial conditions are $N_{B_c}$ = 0, 
and $N_b$ = 1. We use $N_{c_0}$ = 10 and the volume expansion model
for the charm quark density, and a full thermal gluon density.  
The final
$B_c$ population per initial $b\bar b$ pair is  shown in 
Fig.\,\ref{figBcVsT}. The factor of 2 in the axis label 
indicates that on average
an equal number of particle and antiparticle bound states will
be produced. Thus the numbers can be directly compared with the
bound state ratio $10^{-4}\,-\,10^{-5}$ expected if no deconfinement
occurs.   

\begin{figure}[t]
\centerline{  \hspace*{-.cm}
\psfig{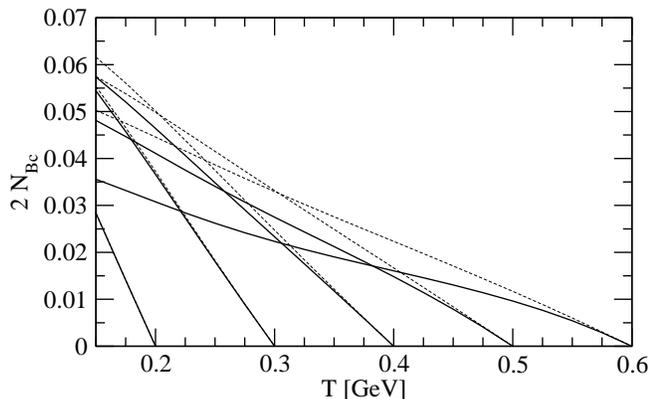}
}
\caption{ \small
 $B_c$ fractional abundance vs. temperature [GeV] for various
 $T_0$ until freeze-out. Dashed line: $\rho_g=\frac{1}{2}\rho_g^{\infty}
$. Other
 parameters same as Fig.\,\ref{figTemp}.
\label{figBcVsT}}
\end{figure}

The nominal initial volume and expansion parameters are used.
 We explore a range $0.2\le T_0\le 0.6$\,GeV of initial temperature  
in steps of 0.1\,GeV. 
We see that the $B_c$ abundance grows
approximately linearly during the entire expansion as temperature decreases. 
This verifies 
that the formation and dissociation reactions do not 
come to equilibrium during the QGP phase, as previously anticipated 
in \cite{TSR99}.
This linearity is not of any inherent physical origin, as can be seen by
changing variables in Eq.\,\ref{eqNBc} from $\tau$ to $T$  
and noting that $N_{B_c} \ll N_b$. The linearity in $T$ 
is seen to follow from the 
numerical values of the product 
$\dot{T}[\tau]\,V[\tau]\,\lambda_{\mathrm{F}}$
remaining almost constant for $T_0 \lesssim 0.6$ GeV . 
The important conclusion we draw from the results shown in  Fig.\,\ref{figBcVsT}
is that we can expect a rather $T_0$ independent 
fractional $B_c$-yield, which for the 
main benchmark of our assumptions is at 5\%.

Shown by dashed lines in Fig.\,\ref{figBcVsT} is the scenario where
gluon density is $1/2$ its thermal
equilibrium value $\rho_g^{\infty}(T)$. This corresponds to an
effective reduction in the available degrees of freedom
for gluons in a QGP for temperatures that are not
yet `asymptotically' large \cite{Blaizot99}. 
One sees the expected
effect of reduced gluon density leading to reduced
dissociation, and hence increased final bound state populations.
However, the increase is substantially less than linear, indicating
that the formation term in the rate equation is dominant.  

In order to better understand this result, and also to 
illustrate the dependence on initial QGP volume and transverse 
velocity, we show  in Fig.\,\ref{figBcVsTo} 
the freeze-out fractional $B_c$-yield per one 
bottom quark pair, as a function 
of $T_0$, while also varying the other parameters.

\begin{figure}[t]
\centerline{  \hspace*{-.cm}
\psfig{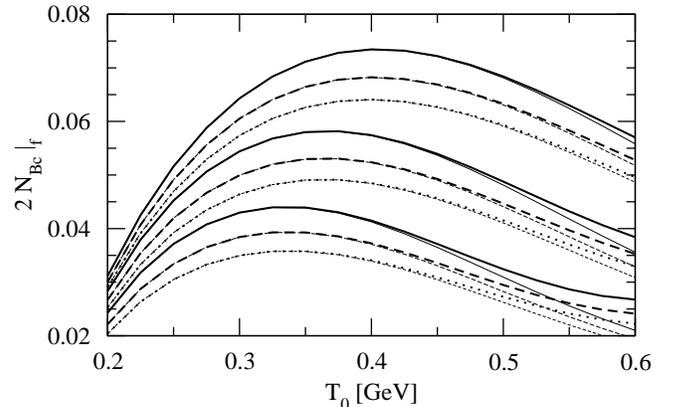}
}
\caption{ \small
 $B_c$ fractional abundance at freeze-out vs. $T_0$
 for $v_r = 0.58c$ (solid), $v_r =0.79c$ (dashed), $v_r =c$ (dotted)
 and $V_0$ = 26\,fm$^3$ (upper 6 lines),
 52\,fm$^3$ (middle 6 lines), 105\,fm$^3$ (lower 6 lines).
 The effect of neglecting additional charm production in the QGP
is shown in the thin lines which are only visible for
$T\geq 0.5$\,GeV. 
\label{figBcVsTo}}
\end{figure}

As $T_0$ increases, the yield initially increases.
However, as the initial temperature further increases, the
dissociation reaction becomes more prominent and adversely impacts the yield,
an effect which is not fully compensated by the additional 
QGP-produced charm available at these
temperatures. This is seen comparing the thick up-curving lines
with thin lines which do not include QGP-produced charm. 
However, this competition between 
formation and dissociation does lead to 
a very broad yield maximum in the vicinity of
the expected range of initial temperatures at RHIC. This effect has
been noted for the nominal expansion and initial volume parameters
in Fig.\,\ref{figBcVsT}.  Here we see this effect persists for
a variety of these parameters.

In order to establish a lower limit for the final $B_c$ 
fractional abundance,
 we show by the dashed and dotted  curves in Fig.\,\ref{figBcVsTo} the
effects of increasing the expansion rate, thus decreasing the
lifetime of the QGP.  We see that this effect is less than that
generated by a change in initial volume, which controls the
formation rate through the change in charm quark density. 
The different initial volumes are controlled by varying $l_i$.
The set of 6 lines in the middle corresponds to $l_i=1$ fm, 
($V_0 = 52$\,fm$^3$), our standard
scenario, while the upper set of lines corresponds to the high density 
case $l_i=0.5$\,fm, ($V_0 = 26$\,fm$^3$), and 
the lower set of lines represents the low density case, 
$l_i=2$\,fm, ($V_0 = 105$\,fm$^3$).

In all of these calculations we have used an initial charm quark number
$N_{c_0}$ = 10, and for the range of temperatures expected at RHIC, this
number will not change appreciably during the QGP lifetime.  
Eq. \ref{eqNBc} then predicts 
that $N_{B_c}$ will be exactly linear with the initial
number of charm quarks, and our numerical results verify this.
The calculated final $B_c$ yields utilizing the average initial
$N_{c_0}$ thus will correctly include the fluctuations 
expected according to a binomial (or Poisson) distribution.
This linear property is not evident in the results shown in
Fig. \ref{figBcVsTo} in terms of the initial charm density, since
there one is changing the density by changing the volume, and this also effects the lifetime of the QGP and through that the final $B_c$ abundance.

Shown in Fig.\,\ref{figBcVsli} is the fractional $B_c$-yield dependence on 
initial volume measured through variation of $l_i$ for 
a range of initial temperatures.  One sees that the temperature variation
is less important than the initial volume effect, at least in the range
of these parameters which we consider. 
For a fixed $T_0$, a change in the initial volume 
not only causes a inverse variation in the initial charm density
$\rho_c \propto l_i^{-1}$, 
but also influences the plasma life time $\tau_L$.  
Empirically it was found that $\tau_L \propto l_i^{1/2}$ is a good fit.
These two factors combine two give the roughly $l_i^{-1/2}$ dependence 
of the final fractional $B_c$ meson yield shown in Fig.\,\ref{figBcVsli}.
\begin{figure}[t]
\centerline{  \hspace*{-.cm}
\psfig{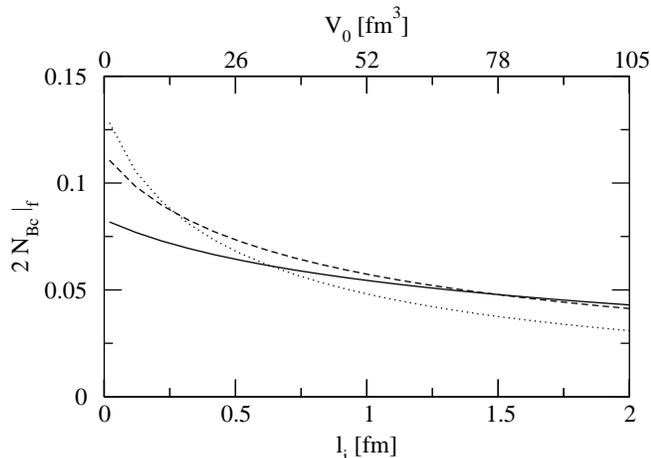}
}
\caption{ \small
 $B_c$ fractional yield at freeze out 
as function of the initial (width)
 parameter $l_i$\,[fm] for $T_0 = 0.5$\,GeV (dotted), 0.4\,GeV
(dashed), and 0.3\,GeV (solid).
\label{figBcVsli}} 
\end{figure}
\subsection{Sensitivity to Breakup and Formation Cross Sections}
The breakup cross section magnitude and shape is an essential part of
our dynamical calculation, since in addition to providing a time scale,
 it
provides through detailed balance the 
relative magnitude of the formation
cross section.  The form in Eq. \ref{eqSigmaB} has been used to 
estimate the breakup rate of $J/\psi$ due to final state collisions
with hadrons, via convolution with the gluon structure function of
the hadron \cite{Kha94}.  Recently, there have been several 
additional attempts to 
model the hadron-quarkonium cross section \cite{NEWSIGMA}, 
which has lead to results which
typically are much larger and have a more rapid rise above threshold.
Here we investigate the sensitivity of our prediction $B_c$ yields to
such variations in the fundamental cross sections.  We show in
Fig. \ref{figsigma} the change in the predicted $B_c$ yields at
RHIC which follow if our breakup cross section in Eq. \ref{eqSigmaB}
is increased by a factor of 2.  Also shown are corresponding results 
if the cross section is assumed to immediately rise to its maximum
value just above threshold, with the same overall magnitudes as
before.  All other model parameters are kept at the nominal
values.  One sees that in all cases the final $B_c$ yields are
increased.  One can understand this behavior as a combination of 
two effects.  First, the detailed balance relation provides the 
relative magnitudes of formation and breakup rates. 
Second, an increase in the magnitude of the cross section 
just decreases the corresponding time scales, allowing the 
favored formation reaction to proceed further toward completion.
Thus the cross section we have 
utilized provides a conservative lower bound 
for our $B_c$ production estimates
at RHIC via this new mechanism.

\begin{figure}[t]
\centerline{  \hspace*{-.cm}
\psfig{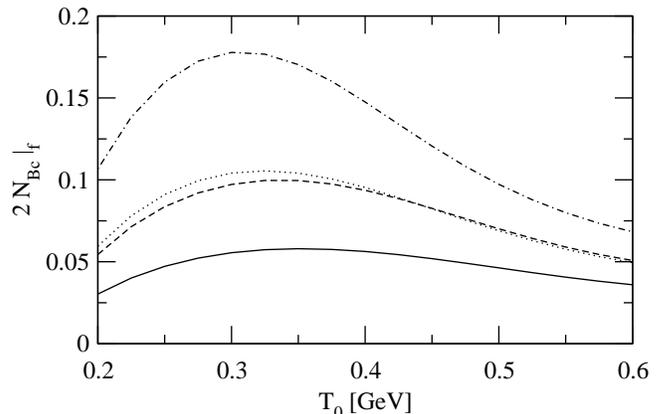}
}
\caption{ \small
 $B_c$ fractional yield at freeze-out
for various theoretical breakup cross sections,
nominal (solid), 2$\times$ nominal (dotted), 
constant (dashed), 2$\times$ constant (dot-dashed).
\label{figsigma}}
\end{figure}

\subsection{Relative excited state $B_c$ yield}\label{2sBcyield}
We have also calculated the ratio of 2S to 1S-state $B_c$ yields
within our model scenario.  This is prompted by the observation
that the corresponding ratio in the charmonium system at SPS 
may serve as a thermometer of the QGP phase \cite{HEINZ97}. 

As a first estimate of this ratio we use as a 2S 
dissociation cross section Eq.\,\ref{eqSigmaB} 
with $\epsilon_{2s}=0.25$ GeV. This of course is only a rough
guess, since one should change the parameters to those 
corresponding to a 2S state, but this state has binding
and energy levels which are very marginal in terms of the
constraints used for the validity of the cross section formula.
The individual population equations can be solved independently,
since both final bound state fractions are small enough 
that the source of b-quarks is not significantly decreased.

Fig.\,\ref{figBc2sBcRatio}
shows the ratio of yields 
$B_c$(2S)/$B_c$(1S) at hadronization as
a function of the  initial temperature for the
three different initial volumes we consider: 
$l_i = 2$\,fm (dotted line), 
 1\,fm (dashed line), and 0.5\,fm (solid line). 
It appears that this ratio is somewhat more sensitive to the initial
temperature than the individual yields, varying by at least a factor
of two within the expected range for RHIC. 
\begin{figure}[t]
\centerline{  \hspace*{-.cm}
\psfig{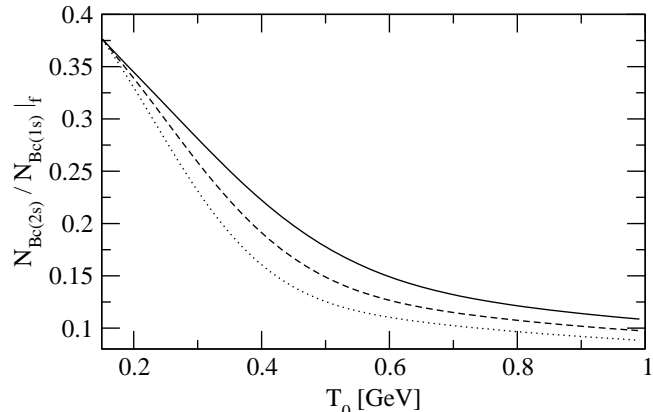}
}
\caption{ \small
 Ratio of $B_c$(2S) to $B_c$(1S) abundance at freeze-out vs.
 initial temperature [GeV] for $l_i = 2$\,fm (dotted  line), 
 1\,fm (dashed  line), and 0.5\,fm (solid  line). 
\label{figBc2sBcRatio}} 
\end{figure}

We also find that this yield ratio is insensitive to
initial charm abundance and production, as charm density 
enters linearly in both the $B_c$(2S) and $B_c$(1S) population equations.
Also, the initial volume as shown in Fig.\,\ref{figBc2sBcRatio}
does not alter the abundance ratio significantly. Thus this yield
ratio may allow one to draw conclusions about initial temperatures 
present in the QGP phase at RHIC, independent of the other parameters. 
To estimate the systematic  uncertainty that would arise 
in such a procedure,  we consider the effects of changing the
cross sections (formation and by detailed balance 
also breakup) in the following cases: (a) increasing the (2S) cross
sections by a factor of two,  (b)
increasing both 2S and 1s cross sections by factor of two,  and
(c) decreasing the $B_c$(1S) and $B_c$(2S) 
binding energies by 100 MeV.
In all three cases the overall shape of the results as 
 shown in Fig.\,\ref{figBc2sBcRatio} remain nearly
unchanged. However, we note  an overall increase 
in the relative yield by a factor of two in case (a),
practically no change in the result in case (b)
and  a decrease by a factor of 1/3 in case (c). 
The ratio $B_c$(2S)/$B_c$(1S) is thus primarily
sensitive to initial temperature, 
but to be able to draw firm conclusions we need 
accurate relative 2S to 1S  cross sections, and 
a good understanding of in-plasma $B_c$ binding energy.

\section{Summary and Conclusion}\label{concl}

We have shown that this new mechanism of quarkonium production 
in a deconfined medium 
predicts the {\it minimum} final state abundance of 
$B_c$ mesons at RHIC to be of the order 
of 5\% per initial $b\bar b$ pair. This result is
relatively insensitive to the initial temperature and volume of the QGP,
as well as to changes in the transverse expansion dynamics. 
There is a linear dependence on the charm quark abundance in the
initial state, which in fact is the primary controlling 
factor in the final state $B_c$ yield. 

Fractional $B_c$ yields at the level of $5 \times 10^{-2}$ significantly
exceed expectations based on initial coherent one step production
 in individual nucleon-nucleon interactions, where
a relative yield in range $10^{-4}\,-\,10^{-5}$ is expected. Such a
small yield would not be observable at RHIC. Even the 500-1000 times
greater
multistep yield we obtain may pose considerable experimental challenges.
However, should such an experiment succeed 
to even roughly confirm these predictions for
the fractional $B_c$ yield, we would have a convincing evidence for the
mobility of charmed quarks over an extended space-time region 
in the dense phase.

While in principle one could argue that incoherent $B_c$ formation 
could also
occur in the hadronic phase in collision of D-mesons with B-mesons, such 
a process requires localization in phase space 
of both these hadrons, which is
highly unlikely. A calculation has been performed for the
analogous case in the $J/\psi$ system, using D-meson interactions 
\cite{REDLICH}.  It was found that this mechanism is negligible 
even at LHC energies, except for possibly some observable effects
in the $\psi^{\prime}$ yield.

We emphasize here that an essential element of this calculation
relies on the assumption that
 colored heavy quarks will be subject to large energy loss
processes in a plasma \cite{energyloss}.
This is necessary if these heavy quarks
are to exist in a common region of the phase space volume.
Without  this stopping,
heavy quarks are highly unlikely to
remain for the required period
of time inside the thermal deconfined phase.
The details of this scenario cannot be fully justified at
present, and considerable effort has to be vested to better understand
the mechanisms which determine the initial phase space distribution
of heavy quarks formed in nuclear collisions.

Another issue is the final state reduction 
of $B_c$ mesons due to collisions with comoving hadrons. 
However, we would expect the smaller size of these bound
states relative to those 
extensively 
studied in the $J/\psi$ system will produce a much less significant
effect in the $B_c$ system. 

Our work relies heavily on the presence of mobile charmed quarks: 
investigation of other bound state formation, such as $J/\psi$,
is currently underway. Initial results indicate that while a significant 
fraction of charm will be in fact bound (at the 10\% 
level) this is too little to significantly alter the 
results we have presented for $B_c$ yields. In fact the 
overall initial charm production uncertainty is at least as
large as this effect. One such source is the possible shadowing
of gluons in nuclei, which could reduce the primary yield by
amounts in the tens of percents \cite{ESKOLA}. 

If the experimental techniques enable observation of these
predicted $B_c$ yields with significant efficiency, one has the 
possibility to embark upon a general
study of both the vacuum and plasma properties of the
$B_c$ system.  Even should this prove not to be possible in the 
RHIC energy range,  it will 
most certainly be easier at the Large Hadron Collider (LHC), 
where the collision energy 
is 30 times greater.  Several bottom quark pairs will be 
produced in each central
nuclear collision, along with literally hundreds of charm quark pairs, 
leading to expectations involving copious production of $B_c$ mesons.
A study of relevant parameters for this production mechanism at LHC is
underway. 

\subsection*{Acknowledgements}
This work was supported in part 
by a grant from the U.S. Department of
Energy,  DE-FG03-95ER40937\,.


\end{narrowtext}
\end{document}